\documentclass[preprintnumbers, floatfix,letterpaper, twocolumn,aps,prd,epsfig]{revtex4-1}
\usepackage{bm,graphicx,dcolumn,epstopdf,epsf, latexsym,mathbbol, amssymb,amsmath,color,slashed, mathrsfs,mathcomp,simplewick}
\usepackage[center]{subfigure}
\usepackage{multirow}
\usepackage{makecell}
\usepackage[colorlinks,linkcolor=blue,citecolor=blue,urlcolor=blue]{hyperref}

\begin{document}
\allowdisplaybreaks
 \newcommand{\bq}{\begin{equation}}
 \newcommand{\eq}{\end{equation}}
 \newcommand{\bqn}{\begin{eqnarray}}
 \newcommand{\eqn}{\end{eqnarray}}
 \newcommand{\ban}{\begin{align}}
 \newcommand{\ean}{\end{align}}
  \newcommand{\nb}{\nonumber}
 \newcommand{\lb}{\label}
 \newcommand{\f}{\frac}
 \newcommand{\p}{\partial}
\newcommand{\PRL}{Phys. Rev. Lett.}
\newcommand{\PLB}{Phys. Lett. B}
\newcommand{\PRD}{Phys. Rev. D}
\newcommand{\CQG}{Class. Quantum Grav.}
\newcommand{\JCAP}{J. Cosmol. Astropart. Phys.}
\newcommand{\JHEP}{J. High. Energy. Phys.}
\newcommand{\NPB}{Nucl. Phys. B}
\newcommand{\Doi}{https://doi.org}
\title{Primordial Spectra of slow-roll inflation at second-order with the Gauss-Bonnet correction}

\author{Qiang Wu${}^{a}$, Tao Zhu${}^{a, b}$, and  Anzhong Wang${}^{a, b}$}
\affiliation{${}^{a}$ Institute for Advanced Physics $\&$ Mathematics, Zhejiang University of Technology, Hangzhou, 310032, China\\
${}^{b}$ GCAP-CASPER, Physics Department, Baylor University, Waco, TX 76798-7316, USA}

\date{\today}

\begin{abstract}

The slow-roll inflation for a single scalar field that couples to the Gauss-Bonnet (GB) term represents an important higher-order
curvature correction inspired by string theory. With the arrival of the era of precision cosmology, it is expected that the high-order
corrections become more and more  important.  In this paper we study the observational predictions of the slow-roll inflation with
the GB term by using the third-order uniform asymptotic approximation method. We calculate explicitly the primordial power spectra,
spectral indices, running of the spectral indices for both scalar and tensor perturbations, and the ratio between tensor and scalar
spectra. These expressions are all written in terms of the Hubble and GB coupling flow parameters and expanded up to the
next-to-leading order in the slow-roll expansions. The upper bounds of errors of the approximations at the third-order are $0.15\%$,
so they represent the most accurate results obtained so far in the literature. We expect that the understanding of the GB corrections
in the primordial spectra and their constraints by forthcoming observational data will provide clues for the UV complete theory of
quantum gravity, such as the string/M-theory.

\end{abstract}


\maketitle

\section{Introduction}
\renewcommand{\theequation}{1.\arabic{equation}} \setcounter{equation}{0}

The inflationary scenario provides a very successful framework for solving the problems with the standard big bang cosmology, as well as accounting for the almost scale-invariant and nearly Gaussian spectra of primordial density perturbations \cite{guth_inflationary_1981, starobinsky_new_1980, sato_first-order_1981, baumann_tasi_2009}. These primordial density perturbations grow to produce the large-scale structure (LSS) seen today in the universe, and meanwhile create the cosmic microwave background (CMB) temperature anisotropy, which has been extensively verified to high precision by WMAP \cite{WMAP2011}, PLANCK \cite{Planck2015-inflation, Planck2015-NG}, and other CMB experiments.

Despite its successes, however, the inflationary scenario also suffers several conceptional  problems, for example, the initial singularity problem \cite{borde_eternal_1994, borde_inflationary_2003} and trans-Planckian problem \cite{martin_trans-planckian_2001, brandenberge_trans-planckian_2013}. All these problems are closely related to the high energy regime that the usual classical general relativity (GR) is known to break down. Because of this, the inflationary scenario in the framework of GR with some corrections could be considered as the effective theory of the complete UV quantum gravity. This has motived a lot of interest to consider quantum gravitational corrections to slow-roll inflation due to higher curvature terms that generically  arise from radiative corrections of quantum gravity, for instance, string/M-theory \cite{BM15} and Horava-Lifshitz  gravity \cite{Wang17}.

One important  higher-order curvature correction is  the GB term coupled with the inflaton. Such a term can be derived from the tree-level effective action of the heterotic string \cite{callan_strings_1985, gross_quartic_1987}, and it has been shown that the theory with such a correction is free of ghost and makes the order of the gravitational equations of motion unchanged. This term could also provide a way to avoid the initial singularity of the Universe \cite{antoniadis_singularity-free_1994, kawai_instability_1998, hawai_evolution_1999, tsujikawa_density_2002, toporensky_nature_2002, tsujikawa_construction_2002}. The GB term has  been extensively studied in the context of various cosmological scenarios, for examples, the accelerating expansion of the universe \cite{bamba_accelerating_2007}, and the slow-roll inflation (see \cite{jiang_inflation_2013,guo_slow-roll_2010, koh_observational_2014, Satoh_slow-roll_2010} and references therein.).

In this work, we are going to particularly focus on the case that the GB term coupled with the slow-roll inflaton field in the early universe and their corrections to the standard slow-roll inflationary perturbations. The general formalism of this theory has already been developed and studied in details in a series papers \cite{jiang_inflation_2013, guo_slow-roll_2010, koh_observational_2014, Satoh_slow-roll_2010,satoh_higher_2008,Chris_HiggsGB_2016,Satoh_circular_2008}. In these works, the primordial perturbation spectra with the GB correction have been investigated and compared with observations. One of the distinguishable features is that it leads to a {\em time-dependent sound speed} associated with the equation of motion for both scalar and tensor perturbations during the slow-roll inflationary period. In the previous works, however, in order to calculate the perturbation spectra, this {\em time-dependent sound speed} has been assumed to be a constant.
In fact, such a treatment is valid only if we calculate the primordial spectra up the first-order in the slow-roll approximation.

Considering the accuracy of current and forthcoming observations, we expect higher-order curvature corrections to the  slow-roll inflation become more and more important. Indeed, as pointed out in \cite{martin_k-inflation_2013, martin_shortcoming_2016, martin_encyclopædia_2014}, to match with the accuracy of the current and forthcoming observations, considerations of the slow-roll approximations beyond the leading-order are highly demanded. Then,
 one needs to consider the {\em time variation} of the sound speed, which is essentially a distinguishable effect of the GB coupling. 

However, considerations of the {\em time variation} of the sound speed make it very difficult to calculate the corresponding power spectra and spectral indices. Recently, we have developed a powerful method, the uniform asymptotic approximation method \cite{zhu_constructing_2014, zhu_inflationary_2014, zhu_high-order_2016}, to calculate precisely the quantum gravitaitonal effects to the power spectra. The robustness of this method has been verified for calculating primordial spectra in $k$-inflation \cite{zhu_power_2014}, and inflation with nonlinear dispersion relations \cite{zhu_gravitational_2014-1, zhu_gravitational_2014-2} and quantum gravitational effects of loop quantum cosmology \cite{zhu_detecting_2015, zhu_scalar_2015, zhu_inflationary_2016}. We note here that this method was first applied to inflationary cosmology in the framewrok of GR  in \cite{habib_inflationary_2002, habib_characterizing_2004, habib_inflationary_2005}, and then we have developed it, so it can be applied to  more general case, including the ones with nonlinear dispersion relations \cite{zhu_constructing_2014, zhu_inflationary_2014, zhu_gravitational_2014-1, zhu_high-order_2016, zhu_gravitational_2014-2, zhu_power_2014, zhu_detecting_2015, zhu_scalar_2015, zhu_inflationary_2016} (For an alternative approach by using Green's function method, see \cite{wei_second-order_2004,Alexandros_Green_2017}). The main purpose of the present paper is to use this powerful method to derive the inflationary observables in slow-roll inflation with the GB correction with high accuracy. With the general expressions of power spectra and spectral indices we obtained in \cite{zhu_power_2014, zhu_gravitational_2014-2}, we calculate explicitly these quantaties for  both scalar and tensor perturbations with the GB correction up to the third-order of the asymptotic uniform approximation. Then,  tensor-to-scalar ratio is also given. These expressions represent a significant improvement over the previous results obtained so far in the literature.

The paper is organized as follows. In Sec. II, we present a brief review of the slow-roll-inflation with the GB coupling, and in Sec. III, we give the most general formulas of the high-order uniform asymptotic approximations. Then,  in Sec. IV, with these general expressions we calculate explicitly the power spectra, spectral indices, and running of the spectral indices of both scalar and tensor perturbations in the slow-roll inflation with the GB correction. Our main conclusions and outlook are summarized in Sec. V.

\section{GB coupled inflation}
\renewcommand{\theequation}{2.\arabic{equation}} \setcounter{equation}{0}

We begin with the action describing the coupling of the inflaton with the GB term,
\bqn\lb{action}
S=\int d^4 x \sqrt{-g}\left[\frac{R}{16 \pi G}-\frac{(\nabla \phi)^2}{2}-V(\phi)-\frac{\xi(\phi)}{2}R^2_{\text{GB}}\right],\nb\\
\eqn
where $\phi$ is the inflation field with a potential $V(\phi)$, $R$ is the Ricci scalar of the spacetime, $R^2_{\text{GB}} \equiv R_{\mu\nu\rho\sigma}R^{\mu\nu\rho\sigma}-4 R_{\mu\nu}R^{\mu\nu}+R^2$ is the GB term, and $\xi(\phi)$ is the GB coupling coefficent. The model is specified by two arbitrary functions, $V(\phi)$ and $\xi(\phi)$.

Now we consider a flat Friedmann-Robertson-Walker (FRW) background,
\bqn
ds^2=-dt^2 +a^2(t) \delta_{ij}dx^{i}dx^{j},
\eqn
where $a(t)$ is the scalar factor of the universe with $t$ being the cosmic time. Then varying the action (\ref{action}) with respect to $g_{\mu\nu}$ and $\phi$ leads to the field equations,
\bqn
&&H^2=\frac{8\pi G}{3}\left(\frac{1}{2}\dot \phi^2+V+12 \dot \xi H^3\right),\lb{friedmann}\\
&&\ddot \phi+3 H \dot \phi+V_{,\phi}+12 \xi_{,\phi} H^2 (\dot H+H^2)=0,\lb{kg_eq}
\eqn
where $H\equiv \dot a/a$ is the Hubble parameter, a dot denotes the derivative with respect to the cosmic time $t$, and $V_{,\phi}=dV(\phi)/d\phi$.
 In order to consider the slow-roll inflation, we need to impose the following slow-roll conditions
\bqn
\frac{1}{2}\dot \phi^2 \ll V,\;\; |\ddot \phi| \ll |3 H \dot \phi|, \;\; |4 \dot \xi H| \ll 1,\;\; |\ddot \xi|\ll |\dot \xi H|.\nb\\
\eqn
With these conditions, it is convenient to introduce two sets of the slow-roll parameters,
the Hubble flow parameters $\epsilon_{n}$ and the flow parameters $\delta_n$ of the GB coupling, which are defined, respectively,  by
\bqn
&&\epsilon_{n+1} \equiv \frac{d\ln \epsilon_{n}}{d\ln a}, \;\;\epsilon_1 \equiv - \frac{\dot H}{H^2},\\
&& \delta_{n+1} \equiv \frac{d\ln \delta_n}{d\ln a},\;\; \delta_1 \equiv \frac{4 \dot \xi H}{M_{\text{Pl}}^2}.
\eqn
In this paper, we also use the conformal time $\eta$ which is defined  as
\bqn
\eta(t)=\int^t_{t_\text{end}}\frac{dt'}{a(t')}.
\eqn
Here $t_{\text{end}}$ denotes the time when the slow-roll inflation ends.

Let us turn to consider the cosmological scalar and tensor perturbations. With the GB coupling, the scalar perturbations obey, \bqn\lb{eom_scalar}
\mu''_{\mathcal{R}}(\eta)+\left(c_{\mathcal{R}}^2 k^2-\frac{z''_{\mathcal{R}}}{z_{\mathcal{R}}}\right)\mu_{\mathcal{R}}(\eta)=0,
\eqn
where a prime denotes a derivative with respect to the conformal time $\eta$, and $\mu_{\mathcal{R}}(\eta)=z_{\mathcal{R}} \mathcal{R}$ is the mode function, $z_{\mathcal{R}}$ and $c_{\mathcal{R}}$ are given by
\bqn
\lb{cs}
c_{\mathcal{R}}^2&=&1+\frac{8\delta \dot \xi H \dot H+2 \delta^2 H^2 (\ddot \xi-\dot \xi H)}{\dot \phi^2 +6 \delta \dot \xi H^3},\\
\lb{zs}
z_{\mathcal{R}}^2 &=& \frac{a^2 (\dot \phi^2 +6 \delta \dot \xi H^3)}{(1-\delta/2)^2 H^2},
\eqn
where $\delta \equiv 4 \dot \xi H M_\text{Pl}^{-2}/(1-4 \dot \xi H/M_\text{Pl}^2)$.

For tensor perturbations, we have
\bqn\lb{eom_tensor}
\mu''_{h}(\eta)+\left(c_{h}^2 k^2-\frac{z''_{h}}{z_h}\right)\mu_{h}(\eta)=0,
\eqn
where   $\mu_{h}(\eta)=z_{h} h_k$ is the mode function, $z_{h}$ and $c_{h}$ are given by
\bqn
\lb{ch}
c_{h}^2&=&1-\frac{ 4(\ddot \xi-\dot \xi H)}{1-4 \dot \xi H},\\
\lb{zh}
z_{h}^2 &=& a^2 (1-4 \dot \xi H).
\eqn

\section{Scalar and Tensor Perturbations with the GB Correction }
\renewcommand{\theequation}{3.\arabic{equation}} \setcounter{equation}{0}

\subsection{General formulas of primordial spectra in the uniform asymptotic approximation}

In this subsection, we present a very brief introduction of the general formulas of primordial perturbations with a slow-varying sound speed.
Details of derivation of these formulas can be found in \cite{zhu_power_2014}.

In the uniform asymptotic approximation, we first write Eqs.(\ref{eom_scalar}) and (\ref{eom_tensor}) in the standard form
\bqn
\frac{d^2 \mu(y)}{dy^2}=\{\lambda^2 \hat g(y)+q(y)\}\mu(y),
\eqn
where we introduce a new variable $y=-k \eta$, $\mu(y)=\mu_{\mathcal{R}}(y)\;\text{and}\;\mu_h(y)$ corresponding to scalar and tensor perturbations respectively, and
\bqn
\lambda^2 \hat g(y)+q(y)=\frac{\nu^2(\eta)-1/4}{y^2}-c^2(\eta),
\eqn
where
\bqn\lb{nu}
\nu^2(\eta) = \eta^2 \frac{z''(\eta)}{z(\eta)} + \frac{1}{4},
\eqn
with $\{\nu(\eta), c(\eta)\} = \{c_{\mathcal{R}}(\eta), c_{\mathcal{R}}(\eta)\}$ and $\{c_h(\eta), c_h(\eta)\}$ corresponding
to scalar and tensor perturbations, respectively. Note that in the above equation, $\lambda$ is supposed to be a large parameter and used to trace the orders of the approximation. In the finial calculation we can set $\lambda=1$ for simplification. Now in order to construct the approximate solutions of the above equation
by using the uniform asymptotic approximation, one needs to choose \cite{zhu_inflationary_2014}
\bqn
q(y)=- \frac{1}{4 y^2},
\eqn
to ensure the convergence of the errors of the approximate solutions. Then,  we have
\bqn
\lambda^2 \hat g(y)=\frac{\nu^2(\eta)}{y^2}-c^2(\eta).
\eqn
Obviously the function $\lambda^2 \hat g(y)$ has a turning point $y_0(\bar \eta_0)=- k \bar \eta_0$, which can be expressed as
\bqn
y_0(\bar \eta_0)=- k \bar \eta_0=\frac{\nu(\bar \eta_0)}{c(\bar \eta_0)}.
\eqn
Then following \cite{zhu_power_2014}, the general formula of the power spectrum reads
\bqn\lb{formula_pw}
\Delta^2(k) &\equiv& \frac{k^3}{2\pi^2} \left|\frac{\mu(y)}{z(\eta)}\right|^2_{y\to 0^+}\nb\\
&=&\frac{k^2}{4\pi^2}\frac{-k \eta}{z^2(\eta)\nu(\eta)}\exp{\left(2 \lambda  \int_y^{\bar y_0} \sqrt{-\hat g(y')}dy' \right)}\nb\\
&&\times \left[1+\frac{\mathscr{H}(+\infty)}{\lambda}+\frac{\mathscr{H}^2(+\infty)}{2 \lambda^2}+\mathcal{O}\left(\frac{1}{\lambda^3}\right)\right].\nb\\
\eqn

In order to calculate the power spectrum in Eq.(\ref{formula_pw}), as we discussed in \cite{zhu_gravitational_2014-2}, in the slow-roll inflation, it is convenient to consider the following expansions,
\bqn
\nu(\eta)&=&\bar \nu_0+\bar \nu_1 \ln{\frac{y}{\bar y_0}}+\frac{1}{2}\bar \nu_2 \ln^2{\frac{y}{\bar y_0}}+\mathcal{O}\left(\ln^3\frac{y}{\bar y_0}\right),\nb\\
c(\eta)&=&\bar c_0+\bar c_1 \ln{\frac{y}{\bar y_0}}+\frac{1}{2}\bar c_2 \ln^2{\frac{y}{\bar y_0}}+\mathcal{O}\left(\ln^3\frac{y}{\bar y_0}\right),\nb\\
\eqn
where $\bar \nu_0\equiv \nu(\bar \eta_0)$, $\bar c_0=c(\bar \eta_0)$, and
\bqn
\bar \nu_1&\equiv&\left. \frac{d \nu(\eta)}{d\ln(-\eta)}\right|_{\eta=\bar \eta_0},\;\;\bar \nu_2\equiv\left. \frac{d^2 \nu(\eta)}{d\ln^2(-\eta)}\right|_{\eta=\bar \eta_0},\nb\\
\bar c_1&\equiv&\left. \frac{d c(\eta)}{d\ln(-\eta)}\right|_{\eta=\bar \eta_0},\;\;\bar c_2\equiv\left. \frac{d^2 c(\eta)}{d\ln^2(-\eta)}\right|_{\eta=\bar \eta_0}.
\eqn
Note that hereafter we use a bar over the quantity to denote that quantity is being evaluated at the turning point $\eta=\bar \eta_0$.

With the above expansions,
the integral $\int \sqrt{\hat g}dy$ can be correspondingly divided into three parts \cite{zhu_power_2014}
\bqn\lb{g3parts}
\int_{y}^{\bar y_0} \sqrt{\hat g(\hat y)}dy=I_1+I_2+I_3,
\eqn
where
\bqn
\lim_{y\to 0}I_1&=& -\bar \nu_0 \left(1+\ln \frac{y}{2\bar y_0}\right),\nb\\
\lim_{y\to 0}I_2&=&\frac{(1-\ln2) \bar c_1 \bar \nu_0}{\bar c_0}-\left(\frac{ \pi^2}{24}-\frac{\ln^22}{2}+\frac{1}{2}\ln^2\frac{y}{\bar y_0}\right)\bar \nu_1,\nb\\
\lim_{y\to 0}I_3&=&- \bar \nu_0 \left(\frac{\pi^2-12\ln^22}{24}\right)\frac{\bar c_1^2}{\bar c_0^2}\nb\\
&&-\bar \nu_0 \left(1-\frac{\pi^2}{24}-\ln2+\frac{\ln^22}{2}\right)\frac{\bar c_2}{\bar c_0}\nb\\
&&+\left(\frac{\zeta(3)}{4}-\frac{\pi^2\ln2}{24}+\frac{\ln^32}{6}-\frac{1}{6}\ln^3\frac{y}{\bar y_0}\right)\bar \nu_2.\nb\\
\eqn

Now, we turn to consider the error control function $\mathscr{H}$, which is given by \cite{zhu_power_2014}
\bqn\lb{Hinfty}
\mathscr{H}(+\infty)&\simeq& \frac{1}{6\bar \nu_0} \left(1+\frac{\bar c_1}{\bar c_0}\right)-\frac{\bar\nu_1(23+12\ln 2)}{72\bar \nu_0^2}\nb\\
&&+\frac{37 \bar c_1^2}{36 \bar c_0^2 \bar \nu _0}-\frac{5 \bar c_2}{36 \bar c_0 \bar \nu _0}-\frac{17 \bar c_1^2 \ln2}{70 \bar c_0^2 \bar \nu _0}+\frac{\bar c_2 \ln2}{6 \bar c_0 \bar \nu _0}.\nb\\
\eqn
Once we get the integral of $\sqrt{g(y)}$ in Eq.(\ref{g3parts}) and error control function in Eq.(\ref{Hinfty}), from Eq.(\ref{formula_pw}) one can easily calculate the power spectra.

Now we turn to consider the corresponding spectral indices. In order to do this, we  first specify the $k$-dependence of $\bar\nu_0(\eta_0)$, $\bar \nu_1(\eta_0)$ through  $\eta_0 = \eta_0(k)$. From the relation $-k\eta_0=\bar \nu_0(\eta_0)/\bar c_0(\eta_0)$, after lengthy technical calculations, we find
\bqn
\frac{d\ln(-\eta_0)}{d\ln k}& \simeq& -1+\frac{\bar c_1}{\bar c_0}-\frac{\bar\nu_1}{\bar\nu_0}-\left(\frac{\bar c_1}{\bar c_0}-\frac{\bar\nu_1}{\bar\nu_0}\right)^2\nb\\
&&+\left(\frac{\bar c_1}{\bar c_0}-\frac{\bar\nu_1}{\bar\nu_0}\right)^3.
\eqn
\begin{widetext}
Then, the spectral indices are given by \cite{zhu_power_2014}
\bqn\lb{index}
n-1&\simeq& \left(3-2 \bar \nu _0\right) + \frac{2 \bar c_1 \bar \nu_0}{\bar c_0} + \left(\frac{1}{6\bar \nu_0^2}-2\ln2\right)\bar \nu_1 + \left(\frac{2\bar  \nu _0 \ln2}{\bar c_0}-\frac{2 \bar \nu _0}{\bar c_0}-\frac{1}{6 \bar c_0 \bar \nu _0}\right)\bar c_2 + \left(\frac{23+12\ln2}{72 \bar \nu _0^2}+\frac{\pi ^2}{12}-\ln^22\right)\bar \nu_2 \nb\\&&
+ \left(\frac{1-12 \bar \nu _0^2 \ln2}{6 \bar c_0^2 \bar \nu _0}\right)\bar c_1^2 +\frac{4\bar \nu_1\bar c_1\ln2}{\bar c_0} +\frac{5-6\ln2}{36\bar c_0\bar \nu_0}\bar c_3 +\left(2-\frac{\pi ^2}{12}+\ln^22-\ln4\right)\frac{\bar \nu_0\bar c_3}{\bar c_0} +\left(\frac{1}{9 }-\frac{\ln2}{35}\right)\frac{17\bar c_1^3}{ \bar c_0^3 \bar \nu _0} \nb\\&&
+\left(2\ln2+2\ln^22-\frac{\pi^2}{6}\right)\frac{\bar \nu_0 \bar c_1^3}{\bar c_0^3 } +\left(\frac{\pi ^2 \bar\nu _0}{4 \bar c_0^2} -\frac{73}{36 \bar c_0^2 \bar \nu _0} -\frac{3 \bar \nu _0 \ln^22}{\bar c_0^2} +\frac{137 \ln2}{210 \bar c_0^2 \bar \nu _0}\right)\bar c_1\bar c_2.
\eqn

Similarly, after some tedious calculations, we find that the running of the spectral index $\alpha \equiv dn/d\ln k$ is given by \cite{zhu_power_2014}
\bqn \lb{running}
\alpha(k)
&\simeq&
\left(\frac{6\bar \nu _0 \ln^22}{\bar c_0^4}-\frac{51 \ln2}{35 \bar c_0^4 \bar \nu _0}+\frac{5\bar \nu _0\ln4 }{\bar c_0^4}+\frac{2 \bar\nu _0}{\bar c_0^4}-\frac{\pi ^2 \bar \nu _0}{2 \bar c_0^4}+\frac{16}{3 \bar c_0^4 \bar \nu _0}\right) \bar c_1^4
+\left(\frac{1}{3 \bar c_0^3 \bar \nu _0}-\frac{2 \bar \nu _0\ln4 }{\bar c_0^3}-\frac{2 \bar \nu _0}{\bar c_0^3}\right) \bar c_1^3
\nb\\&&
+\left(\frac{58 \ln 2}{21 \bar c_0^3 \bar \nu _0}+\frac{\pi ^2 \bar \nu _0}{\bar c_0^3}-\frac{83}{9 \bar c_0^3 \bar \nu _0}-\frac{12 \bar \nu _0 \ln^22}{\bar c_0^3}-\frac{12 \bar \nu _0 \ln2}{\bar c_0^3}\right) \bar c_2 \bar c_1^2
+\frac{2 \bar \nu _0 \bar c_1^2}{\bar c_0^2}
+\left(\frac{6 \ln2}{\bar c_0^2}+\frac{6}{\bar c_0^2}+\frac{1}{6 \bar c_0^2 \bar \nu _0^2}\right) \bar \nu _1 \bar c_1^2
\nb\\&&
+\left(\frac{6 \bar \nu _0\ln2 }{\bar c_0^2}-\frac{1}{2 \bar c_0^2 \bar \nu _0}\right) \bar c_2 \bar c_1
+\left(\frac{4 \bar \nu _0 \ln^22}{\bar c_0^2}-\frac{86 \ln2}{105 \bar c_0^2 \bar \nu _0}-\frac{\pi ^2 \bar \nu _0}{3 \bar c_0^2}+\frac{2}{\bar c_0^2 \bar \nu _0}\right) \bar c_3 \bar c_1
+\left(\frac{1}{6 \bar c_0 \bar \nu _0^2}-\frac{6 \ln2}{\bar c_0}\right) \bar \nu _2 \bar c_1
\nb\\&&
-\frac{4 \bar \nu _1 \bar c_1}{\bar c_0}
+\left(\frac{3 \bar \nu _0 \ln^22}{\bar c_0^2}-\frac{137 \ln2}{210 \bar c_0^2 \bar \nu _0}-\frac{\pi ^2 \bar \nu _0}{4 \bar c_0^2}+\frac{73}{36 \bar c_0^2 \bar \nu _0}\right) \bar c_2^2
+\left(\frac{1}{3 \bar \nu _0^3}+\frac{2}{\bar \nu _0}\right) \bar \nu _1^2
+\left(\frac{2 \bar \nu _0}{\bar c_0}+\frac{1}{6 \bar c_0 \bar \nu _0}-\frac{2\bar \nu _0 \ln2 }{\bar c_0}\right) \bar c_3
\nb\\&&
+\frac{\pi ^2 \bar c_4 \bar \nu _0}{12 \bar c_0}
+\left(-\frac{6 \ln2}{\bar c_0}-\frac{1}{6 \bar c_0 \bar \nu _0^2}\right)\bar c_2 \bar \nu_1
+2 \bar \nu _1
+\left(\ln4-\frac{1}{6 \bar \nu _0^2}\right) \bar \nu _2
+\left(\ln^22-\frac{\pi ^2}{12}-\frac{\ln2}{6 \bar \nu _0^2}-\frac{23}{72 \bar \nu _0^2}\right) \bar \nu _3
\nb\\&&
+\frac{\bar c_4 \bar \nu _0\ln4 }{\bar c_0}
-\frac{2 \bar c_2 \bar \nu _0}{\bar c_0}-\frac{\bar \nu _0\bar c_4 \ln^22 }{\bar c_0}-\frac{2 \bar c_4 \bar \nu _0}{\bar c_0}-\frac{5 \bar c_4}{36 \bar c_0 \bar \nu _0}+\frac{\ln2 \bar c_4}{6 \bar c_0 \bar \nu _0}.
\eqn
\end{widetext}

In the above, we present all the formulas (Eqs.(\ref{formula_pw}), (\ref{index}), and (\ref{running})) that can be directly used to calculate the primordial perturbation spectra from different inflation models. Note
 that these formulas are easy to use because they only depend on the quantities $H(\eta)$, ($c_0, c_1, c_2$), ($\nu_0, \nu_1, \nu_2$) evaluated at the turning point. These quantities can be easily calculated from Eqs.(\ref{cs}, \ref{zs}) for scalar perturbations, Eqs.(\ref{ch}, \ref{zh}) for tensor perturbations. In the following subsections, we apply these formulas to calculate the slow-roll power spectra with the GB correction for both scalar and tensor perturbations.

\subsection{Scalar Spectrum}

We first consider the scalar perturbations. As we already pointed out in the introduction, in order to match the accurchy of forthcoming observations, we need to calculate the primordial spectra up to the next-to-leading order (second-order) in the expansions of the slow-roll approximation. For this purpose, we only need to consider the quantities $(c_{\mathcal{R} 0}, \; c_{\mathcal{R} 1}, \; c_{\mathcal{R} 2})$ and $(\nu_{\mathcal{R} 0}, \; \nu_{\mathcal{R} 1}, \; \nu_{\mathcal{R} 2})$ up to the second-order in the slow-roll expansions.

For the slow-varying sound speed $c_{\mathcal{R}}$, from Eq.(\ref{cs}) we find
\bqn
\bar c_{\mathcal{R}0}&=&1+\frac{\bar \delta _1^3+4 \bar \delta _1^2 \bar \epsilon _1}{4 \left(\bar \delta _1-2 \bar \epsilon _1\right)}\nb\\
&&+\frac{7 \bar \delta _1^5+8 \bar \delta _1^4 \bar \epsilon _1+12 \bar \delta _1^3 \bar \delta _2 \bar \epsilon _1-20 \bar \delta _1^3 \bar \epsilon _1^2}{8 \left(\bar \delta _1-2 \bar \epsilon _1\right){}^2} +\mathcal{O}(\bar \epsilon_i^4),\nb\\
\bar c_{\mathcal{R} 1} &\equiv&  \frac{d c_{\mathcal{R}}}{d\ln (-\eta)} =\mathcal{O}(\bar \epsilon_i^3),\nb\\
 \bar c_{\mathcal{R} 2} &\equiv& \frac{d^2 c_{\mathcal{R}}}{d\ln^2 (-\eta)} = \mathcal{O}(\bar \epsilon_i^4).
\eqn
For $\nu_{\mathcal{R}}$, from Eqs.(\ref{nu}) and (\ref{zs}) we find
\bqn
\bar \nu_{\mathcal{R} 0}&=&\frac{3}{2}+\frac{\bar \delta _1 \bar \delta _2+2 \bar \delta _1 \bar \epsilon _1-4 \bar \epsilon _1^2-2 \bar \epsilon _1 \bar \epsilon_2}{2 \left(\bar \delta _1-2 \bar \epsilon _1\right)}\nb\\
&&+\frac{1}{12 \left(\bar \delta _1-2 \bar \epsilon _1\right){}^2} \Big(-3 \bar \delta _1^3 \bar \delta _2-4 \bar \delta _1^2 \bar \delta _2 \bar \delta _3+18 \bar \delta _1^2 \bar \delta _2 \bar \epsilon _1\nb\\
&&~~~~~~~~~~~~~~~~~~+8 \bar \delta _1 \bar \delta _2^2 \bar \epsilon _1+8 \bar \delta _1 \bar \delta _2 \bar \delta _3 \bar \epsilon _1+12 \bar \delta _1^2 \bar \epsilon _1^2\nb\\
&&~~~~~~~~~~~~~~~~~~-48 \bar \delta _1 \bar \epsilon _1^3+48 \bar \epsilon _1^4+4 \bar \delta _1^2 \bar \epsilon _1 \bar \epsilon _2-4 \bar \delta _1 \bar \delta _2 \bar \epsilon _1 \bar \epsilon _2\nb\\
&&~~~~~~~~~~~~~~~~~~~-76 \bar \delta _1 \bar \epsilon _1^2 \bar \epsilon _2+88 \bar \epsilon _1^3 \bar \epsilon _2-4 \bar \delta _1 \bar \epsilon _1 \bar \epsilon _2^2\nb\\
&&~~~~~~~~~~~~~~~~~~~-4 \bar \delta _1 \bar \epsilon _1 \bar \epsilon _2 \bar \epsilon _3+8 \bar \epsilon _1^2 \bar \epsilon _2 \bar \epsilon _3\Big)+\mathcal{O}(\bar \epsilon_i^3),\nb\\
\eqn
\bqn
 \bar \nu_{\mathcal{R}1} & \equiv& \frac{d \nu_{\mathcal{R}}}{d\ln (-\eta)} \nb\\
 &=&\frac{1}{2(\bar \delta_1-2\bar \epsilon_1)^2} \Big(-\bar \delta _1^2 \bar \delta _2 \bar \delta _3+2 \bar \delta _1 \bar \delta _2^2 \bar \epsilon _1+2 \bar \delta _1 \bar \delta _2 \bar \delta _3 \bar \epsilon _1\nb\\
 &&~~~~~~~~~~~~~~~~~~ -2 \bar \delta _1^2 \bar \epsilon _1 \bar \epsilon _2-4 \bar \delta _1 \bar \delta _2 \bar \epsilon _1 \bar \epsilon _2+8 \bar \delta _1 \bar \epsilon _1^2 \bar \epsilon _2\nb\\
 &&~~~~~~~~~~~~~~~~~~ -8 \bar \epsilon _1^3 \bar \epsilon_2+2 \bar \delta _1 \bar \epsilon _1 \bar \epsilon _2^2\nb\\
 &&~~~~~~~~~~~~~~~~~~ +2 \bar \delta _1 \bar \epsilon _1 \bar \epsilon _2 \bar \epsilon_3-4 \bar \epsilon _1^2 \bar \epsilon _2 \bar \epsilon _3\Big)+\mathcal{O}(\bar \epsilon_i^3),\nb\\
 \eqn
 and
 \bqn
 \bar \nu_{\mathcal{R}1} \equiv \frac{d^2 \nu_{\mathcal{R}}}{d\ln^2 (-\eta)} &=& \mathcal{O}(\bar \epsilon_i^3).
\eqn

Then, using the above expansions, the power spectrum for the curvature perturbation $\mathcal{R}$ can be calculated via Eq.(\ref{formula_pw}). After tedious calculations we obtain,
\begin{widetext}
\bqn
\Delta_{\mathcal{R}}^2(k) &=&\frac{181 \bar H^2}{36 e^3 \pi ^2 \left(2   \bar \epsilon _1-  \bar  \delta _1\right) M_{\text{Pl}}^2} \Bigg\{1+2\ln2\bar \epsilon_1+\frac{1}{2 \bar \epsilon _1-\bar \delta _1} \Big[\bar \delta _1 \bar \epsilon _1 \frac{315}{181}-\frac{\bar \delta_1^2}{2}+\bar \delta _2 \bar \delta _1 \left(-\frac{114}{181}-\ln {2}\right)\nb\\
&&-\frac{992}{181}\bar \epsilon _1^2 +\bar \epsilon _1 \bar \epsilon _2 \left(2 \ln {2}-\frac{134}{181}\right)\Big]
+\frac{1}{\left(2 \bar \epsilon _1-\bar \delta _1\right){}^2} \Big[\bar \delta _2 \bar \delta _1^2 \bar \epsilon _1 \left(-\frac{652}{181}+2\ln ^2{2}+\frac{456 \ln {2}}{181}\right)\nb\\
&&+\bar \delta _1^2 \bar \epsilon _1 \bar \epsilon _2
   \left(-\frac{2827}{1629}+\frac{\pi ^2}{12}-\ln ^2{2}-\frac{47 \ln {2}}{181}\right)+\bar \delta _1 \bar \epsilon _1^3
   \left(-\frac{180}{181}-8 \ln ^2{2}+\frac{1796 \ln {2}}{181}\right)\nb\\
  &&+\bar \delta _2 \bar \delta _1 \bar \epsilon _1^2
   \left(\frac{768}{181}-4 \ln ^2{2}+\frac{536 \ln {2}}{181}\right)+\bar \delta _1 \bar \epsilon _1 \bar \epsilon _2^2
   \left(\frac{172}{1629}-\frac{\pi ^2}{12}+\ln ^2{2}-\frac{134 \ln {2}}{181}\right)\nb\\
 &&+\bar \delta _2^2 \bar \delta _1
   \bar \epsilon _1 \left(-\frac{1034}{1629}-\frac{\pi ^2}{12}+\ln ^2{2}+\frac{228 \ln {2}}{181}\right)+\bar \delta _2
   \bar \delta _3 \bar \delta _1 \bar \epsilon _1 \left(-\frac{1034}{1629}-\frac{\pi ^2}{12}+\ln ^2{2}+\frac{228 \ln
   {2}}{181}\right)\nb\\
   &&+\bar \delta _2 \bar \delta _1 \bar \epsilon _1 \bar \epsilon _2 \left(\frac{2266}{1629}+\frac{\pi ^2}{6}-4 \ln
   ^2{2}-\frac{188 \ln {2}}{181}\right)+\bar \delta _1 \bar \epsilon _1 \bar \epsilon _2 \bar \epsilon _3
   \left(\frac{172}{1629}-\frac{\pi ^2}{12}+\ln ^2{2}-\frac{134 \ln {2}}{181}\right)\nb\\
   &&+\bar \delta _1^3 \bar \epsilon _1
   \left(\ln {2}-\frac{677}{362}\right)+\bar \delta _1^2 \bar \epsilon _1^2 \left(\frac{655}{181}-\frac{ 630\ln
   2}{181}+2\ln^2 {2} \right)+\bar \delta _1 \bar \epsilon _1^2 \bar \epsilon _2 \left(\frac{18940}{1629}-\frac{\pi
   ^2}{3}-8 \ln {2}\right)\nb\\
   &&+\frac{3 \bar \delta _1^4}{4}+2 \bar \delta _2 \bar \delta _1^3+\bar \delta _2^2 \bar \delta _1^2
   \left(\frac{217}{362}+\frac{\ln ^2{2}}{2}+\frac{114 \ln {2}}{181}\right)+\bar \delta _2 \bar \delta _3 \bar \delta _1^2
   \left(\frac{517}{1629}+\frac{\pi ^2}{24}-\frac{\ln ^2{2}}{2}-\frac{114 \ln {2}}{181}\right)\nb\\
   &&+\bar \epsilon _1^2
   \bar \epsilon _2^2 \left(-\frac{22}{181}+2 \ln ^2{2}-\frac{268 \ln {2}}{181}\right)+\bar \epsilon _1^3 \bar \epsilon _2
   \left(-\frac{16924}{1629}+\frac{\pi ^2}{3}+4 \ln ^2{2}+\frac{188 \ln {2}}{181}\right)\nb\\
   &&+\bar \epsilon _1^2 \bar \epsilon
   _2 \bar \epsilon _3 \left(-\frac{344}{1629}+\frac{\pi ^2}{6}-2 \ln ^2{2}+\frac{268 \ln {2}}{181}\right)+\bar \epsilon
   _1^4 \left(\frac{1172}{181}-\frac{2520 \ln 2}{181}+8 \ln ^2 {2} \right)\Big]\nb\\
   &&+ \mathcal{O}(\bar  \epsilon_i^3)\Bigg\}.
\eqn
Similarly,  the spectral index of scalar spectrum reads
\bqn
n_s-1&\simeq& -2\bar \epsilon_1+\frac{1}{2 \bar \epsilon _1-\bar  \delta _1} \left(\bar  \delta _1 \bar \delta _2-2 \bar \epsilon _2
\bar \epsilon _1\right)+\frac{1}{\left(2 \bar \epsilon _1-\bar \delta _1\right){}^2} \Big[\frac{1}{2} \bar \delta _1^3 \bar \delta _2+\left(\frac{17}{27}+\ln {2}\right) \bar \delta _1^2 \bar \delta _2 \bar \delta _3-3
   \bar \delta _1^2 \bar \delta _2 \bar \epsilon _1 \nb \\
   &&+\left(-\frac{34}{27}-2 \ln {2}\right) \bar \delta _1 \bar \delta _2^2
   \bar \epsilon _1+\left(-\frac{34}{27}-2 \ln {2}\right)-2 \bar \delta _1^2 \bar \epsilon _1^2+8 \bar \delta _1 \bar \epsilon
   _1^3-8 \bar \epsilon _1^4+\left(-\frac{20}{27}+2 \ln {2}\right) \bar \delta _1^2 \bar \epsilon _1 \bar \epsilon
   _2 \nb \\
   &&+\left(\frac{14}{27}+4 \ln {2}\right) \bar \delta _1 \bar \delta _2 \bar \epsilon _1 \bar \epsilon
   _2+\left(\frac{350}{27}-8 \ln {2}\right) \bar \delta _1 \bar \epsilon _1^2 \bar \epsilon
   _2+\left(-\frac{404}{27}+8 \ln {2}\right) \bar \epsilon _1^3 \bar \epsilon _2+\left(\frac{20}{27}-2 \ln
   {2}\right) \bar \delta _1 \bar \epsilon _1 \bar \epsilon _2^2 \nb \\
   &&+\left(\frac{20}{27}-2 \ln {2}\right) \bar \delta _1
   \bar \epsilon _1 \bar \epsilon _2 \bar \epsilon _3+\left(-\frac{40}{27}+4 \ln {2}\right) \bar \epsilon _1^2 \bar \epsilon _2
   \bar \epsilon _3\Big]+ \mathcal{O}(\bar \epsilon_i^3),
\eqn
and the running of the scalar spectral index is expressed as
\bqn
\alpha_s \simeq \frac{1}{\left(2 \bar \epsilon _1-\bar \delta _1\right){}^2} &&\Big(8 \bar \delta _1 \bar \epsilon _2 \bar \epsilon _1^2+2 \bar \delta _1 \bar \delta _2^2 \bar \epsilon _1+2 \bar \delta _1 \bar \epsilon _2^2 \bar \epsilon _1+2 \bar \delta _1 \bar \delta _2 \bar \delta _3 \bar \epsilon _1-2 \bar \delta _1^2 \bar \epsilon _2 \bar \epsilon _1-4 \bar \delta _1 \bar \delta _2 \bar \epsilon _2 \bar \epsilon _1+2 \bar \delta _1 \bar \epsilon _2 \bar \epsilon _3 \bar \epsilon _1\nb \\&&-\bar \delta _1^2 \bar \delta _2 \bar \delta _3-8 \bar \epsilon _2 \bar \epsilon _1^3-4 \bar \epsilon _2 \bar \epsilon _3 \bar \epsilon _1^2\Big)+\mathcal{O}( \bar \epsilon_i^3).
\eqn
\end{widetext}

\subsection{Tensor Spectrum}

Now we consider the tensor spectrum. First we need to derive the expressions of $ \nu_{h 0}, \nu_{ h 1}, \nu_{h 2}$, and $c_{h 0}, c_{ h 1}, c_{ h 2}$. Repeating similar calculations for scalar perturbations, we obtain
\bqn
\bar \nu_{h 0} &=& \frac{3}{2}+\bar \epsilon _1-\frac{\bar \delta _1 \bar \delta _2}{2}+\bar \epsilon _1^2+\frac{4 \bar \epsilon _1 \bar \epsilon _2}{3}+\mathcal{O}(\bar \epsilon^3),~~~~~~~ \\
\bar \nu_{h1}&\equiv & \frac{d\nu_h}{d\ln(-\eta)} = -\bar \epsilon_1 \bar \epsilon_2 +\mathcal{O}(\bar \epsilon^3),\\
\bar \nu_{h 2} &\equiv & \frac{d^2\nu_h}{d\ln^2(-\eta)} =\mathcal{O}(\bar \epsilon^3),
\eqn
and
\bqn
\bar c_{h0}&=&1+\frac{\bar \delta _1}{2}+\frac{3 \bar \delta _1^2}{8}-\frac{\bar \delta _1 \bar \delta _2}{2}-\frac{\bar \delta_1 \bar \epsilon _1}{2} +\mathcal{O}(\bar \epsilon_i^3),~~~~~~~~ \\
\bar c_{ h1} & \equiv & \frac{d c_{h}}{d\ln (-\eta)} = -\frac{1}{2}\bar \delta_1 \bar \bar \delta_2 +\mathcal{O}(\bar \epsilon_i^3),\\
 \bar c_{h 2} & \equiv & \frac{d^2 c_{h}}{d\ln^2 (-\eta)} =\mathcal{O}(\bar \epsilon_i^3).
\eqn

Then,  the power spectrum for the tensor perturbation $h_k$ reads
\bqn
\Delta_{h}^2(k) &=&\frac{181 \bar H^2}{36 e^3 \pi ^2} \Big[1+2\ln2 \bar \epsilon _1-\frac{\bar \delta _1}{2}-\frac{\bar \delta_1^2}{8}+\frac{67 \bar \delta _1 \bar \delta _2}{181} \nb\\
&&~~~~~~~~~~~~ +\frac{1}{2} \ln {2} \bar \delta _1 \bar \delta _2-\frac{496 \bar \epsilon _1}{181}+\frac{1039 \bar \delta _1 \bar \epsilon _1}{362} \nb\\
&&~~~~~~~~~~~~ -\ln {2} \bar \delta _1 \bar \epsilon _1+\frac{293 \bar \epsilon _1^2}{181}-\frac{315}{181} 2\ln2 \bar \epsilon _1^2\nb\\
&&~~~~~~~~~~~~ +\ln {2} 2\ln2 \bar \epsilon _1^2-\frac{4636 \bar \epsilon _1 \bar \epsilon _2}{1629}+\frac{1}{12} \pi ^2 \bar \epsilon _1 \bar \epsilon _2\nb\\
&&~~~~~~~~~~~~ +\frac{496}{181} \ln {2} \bar \epsilon _1 \bar \epsilon _2-\ln ^2{2} \bar \epsilon _1 \bar \epsilon _2 +\mathcal{O}( \bar \epsilon_i^3)\Big], \nb\\
\eqn
while the spectral index is given by
\bqn
n_t &\simeq& -2 \bar \epsilon _1-\frac{1}{2} \bar  \delta _1 \bar  \delta _2-2 \bar \epsilon_1^2\nb\\
 && +\left(-\frac{74}{27}+2\ln 2\right) \bar \epsilon _1 \bar \epsilon _2+\mathcal{O}( \bar \epsilon_i^3),
\eqn
and the running of the tensor spectral index is expressed as
\bqn
\alpha_t\simeq-2 \bar \epsilon _1 \bar \epsilon _2+\mathcal{O}( \bar \epsilon_i^3).
\eqn

\subsection{Expressions at Horizon Crossing}

In the last two subsections, we have obtained the expressions of the power spectra, spectral indices, and running of spectral indices for both scalar and tensor perturbations. It should be noted that all these expressions were evaluated at the turning point $y = y_0$. However, usually those expressions were expressed in terms of the slow-roll parameters which are evaluated at the time $\eta_\star$ when scalar or tensor perturbation modes cross the horizon, i.e., $a(\eta_\star) H (\eta_\star) = c_s(\eta_\star) k$ for scalar perturbations and $a(\eta_\star) H (\eta_\star) = c_h(\eta_\star) k$ for tensor perturbations. Consider modes with the same wave number $k$, it is easy to see that the scalar and tensor modes left the horizon at different times if $c_s(\eta) \neq c_h(\eta)$.  When $c_s(\eta_\star) > c_h(\eta_\star)$, the scalar mode leaves the horizon later than the tensor mode, and for $c_s(\eta_\star) < c_h(\eta_\star)$, the scalar mode leaves the horizon before the tensor one.

In this case, caution must be taken for the evaluation time for all the inflationary observables. As we have two different horizon crossing times, it is reasonable to rewrite all our results in terms of quantities evaluated at the later time, i.e., we should evaluate all expressions at scalar horizon crossing time $a(\eta_\star) H (\eta_\star) = c_s(\eta_\star) k$ for $c_s(\eta_\star) > c_h(\eta_\star)$ and at tensor horizon crossing $a(\eta_\star) H (\eta_\star) = c_h(\eta_\star) k$ for $c_s(\eta_\star) <c_h(\eta_\star)$. In the following, we present all the expressions for both cases, separately.

\subsubsection{$c_s(\eta_\star)>c_t(\eta_\star)$}

For $c_s(\eta_\star)>c_t(\eta_\star)$, as the scalar mode leaves the horizon later than the tensor mode, we shall rewrite all the expressions in terms of quantities evaluated at the time when the scalar leaves the Hubble horizon at
$a(\eta_\star) H (\eta_\star) = c_s(\eta_\star) k$. Skipping all the tedious calculations, we find that the scalar spectrum can be written finally in the form
\begin{widetext}
\bqn
\Delta_{\mathcal{R}}^2(k) &=&\frac{181 H_{\star}^2}{36 e^3 \pi^2 (2 {\epsilon_\star} _1-{\delta_\star}_1)}\Big\{ 1+2\ln3{\epsilon_\star} _1+ \frac{1}{(2 {\epsilon_\star} _1-{\delta_\star} _1)}\Big[-\frac{{\delta_\star} _1^2}{2}-\left(\frac{114}{181}+\ln3\right) {\delta_\star} _1 {\delta_\star}
   _2\nb\\
   && +\frac{315}{181} {\delta_\star} _1 {\epsilon_\star}
   _1-\frac{992}{181} {\epsilon_\star}
   _1^2-\left(\frac{134}{181}-2 \ln3\right) {\epsilon_\star} _1 {\epsilon_\star} _2\Big]+\frac{1}{\left(2 {\epsilon_\star} _1-{\delta_\star} _1\right){}^2}\Big[ {\delta_\star} _1^2 {\epsilon_\star} _1^2 \left(\frac{1603}{543}+2 \ln^2{3}-\frac{630 \ln3}{181}\right)\nb \\
   &&+{\delta_\star} _2 {\delta_\star} _1^2 {\epsilon_\star} _1 \left(-\frac{1775}{543}+2 \ln^2{3}+\frac{456 \ln3}{181}\right)+{\delta_\star} _1^2 {\epsilon_\star} _1 {\epsilon_\star} _2 \left(\frac{\pi ^2}{12}-\frac{2827}{1629}-\ln^2{3}-\frac{47 \ln3}{181}\right)\nb \\
   &&+{\delta_\star} _1 {\epsilon_\star} _1^3 \left(\frac{908}{543}-8 \ln^2{3}+\frac{1796 \ln3}{181}\right)+{\delta_\star} _2 {\delta_\star} _1 {\epsilon_\star} _1^2 \left(\frac{1942}{543}-4 \ln^2{3}+\frac{536 \ln3}{181}\right)\nb \\
   &&+{\delta_\star} _1 {\epsilon_\star} _1 {\epsilon_\star} _2^2 \left(-\frac{\pi ^2}{12}+\frac{172}{1629}+\ln^2{3}-\frac{134 \ln3}{181}\right)+{\delta_\star} _2^2 {\delta_\star} _1 {\epsilon_\star} _1 \left(-\frac{\pi ^2}{12}-\frac{1034}{1629}+\ln^2{3}+\frac{228 \ln3}{181}\right)\nb \\
   &&+{\delta_\star} _2 {\delta_\star} _3 {\delta_\star} _1 {\epsilon_\star} _1 \left(-\frac{\pi ^2}{12}-\frac{1034}{1629}+\ln^2{3}+\frac{228 \ln3}{181}\right)+{\delta_\star} _2 {\delta_\star} _1 {\epsilon_\star} _1 {\epsilon_\star} _2 \left(\frac{\pi ^2}{6}+\frac{94}{1629}-4 \ln^2{3}-\frac{188 \ln3}{181}\right)\nb \\
   &&+{\delta_\star} _1 {\epsilon_\star} _1 {\epsilon_\star} _2 {\epsilon_\star} _3 \left(-\frac{\pi ^2}{12}+\frac{172}{1629}+\ln^2{3}-\frac{134 \ln3}{181}\right)\nb\\
   && +{\delta_\star} _1^3 {\epsilon_\star} _1 \left(\ln3-\frac{677}{362}\right)+{\delta_\star} _1 {\epsilon_\star} _1^2 {\epsilon_\star} _2 \left(-\frac{\pi ^2}{3}+\frac{17854}{1629}-8 \ln3\right)\nb \\
   &&+\frac{3 {\delta_\star} _1^4}{4}+2 {\delta_\star} _2 {\delta_\star} _1^3+{\delta_\star} _2^2 {\delta_\star} _1^2 \left(\frac{1013}{1086}+\frac{\ln^2{3}}{2}+\frac{114 \ln3}{181}\right)+{\delta_\star} _2 {\delta_\star} _3 {\delta_\star} _1^2 \left(\frac{\pi ^2}{24}+\frac{517}{1629}-\frac{1}{2} \ln^2{3}-\frac{114 \ln3}{181}\right)\nb \\
   &&+{\epsilon_\star} _1^4 \left(\frac{2068}{543}+8 \ln^2{3}-\frac{2520 \ln3}{181}\right)+{\epsilon_\star} _1^2 {\epsilon_\star} _2^2 \left(\frac{658}{543}+2 \ln^2{3}-\frac{268 \ln3}{181}\right)\nb \\
   &&+{\epsilon_\star} _1^3 {\epsilon_\star} _2 \left(\frac{\pi ^2}{3}-\frac{14752}{1629}+4 \ln^2{3}+\frac{188 \ln3}{181}\right)\nb\\
   && +{\epsilon_\star} _1^2 {\epsilon_\star} _2 {\epsilon_\star} _3 \left(\frac{\pi ^2}{6}-\frac{344}{1629}-2 \ln^2{3}+\frac{268 \ln3}{181}\right)\Big]+\mathcal{O}({\epsilon_\star}_i^3).
   \Big\}
\eqn
The spectral index $n_s$ is given by
 \bqn
 n_{s}-1&\simeq& -2{\epsilon_\star}_1+\frac{1}{2 {\epsilon_\star} _1-{\delta_\star} _1} \left({\delta_\star} _1 {\delta_\star} _2-2 {\epsilon_\star} _2 {\epsilon_\star} _1\right)
  +\frac{1}{\left(2 {\epsilon_\star} _1-{\delta_\star} _1\right){}^2}\Big\{8 {\delta_\star} _1 {\epsilon_\star}_1^3\nb\\
 && -2 {\delta_\star} _1^2 {\epsilon_\star} _1^2-3 {\delta_\star} _1^2 {\delta_\star} _2 {\epsilon_\star} _1+{\delta_\star} _1 {\epsilon_\star} _2 {\epsilon_\star} _1^2 \left(\frac{350}{27}-8\ln3\right)\nb\\
 &&+{\delta_\star} _1 {\delta_\star} _2^2 {\epsilon_\star} _1 \left(-\frac{34}{27}-2 \ln3\right)+{\delta_\star} _1 {\epsilon_\star} _2^2 {\epsilon_\star} _1 \left(\frac{20}{27}-2 \ln3\right)\nb\\
 && +{\delta_\star} _1 {\delta_\star} _2 {\delta_\star} _3
  {\epsilon_\star} _1\left(-\frac{34}{27}-2 \ln3\right)+{\delta_\star} _1^2 {\epsilon_\star} _2 {\epsilon_\star} _1\left(2 \ln3-\frac{20}{27}\right)\nb\\
 &&+{\delta_\star} _1{\delta_\star} _2 {\epsilon_\star} _2{\epsilon_\star} _1 \left(\frac{14}{27}+4 \ln3\right)+{\delta_\star} _1 {\epsilon_\star} _2{\epsilon_\star} _3 {\epsilon_\star} _1 \left(\frac{20}{27}-2 \ln3\right)\nb\\
 && +\frac{1}{2}
   {\delta_\star} _1^3 {\delta_\star} _2+{\delta_\star} _1^2 {\delta_\star} _2 {\delta_\star} _3\left(\frac{17}{27}+\ln3\right)-8 {\epsilon_\star} _1^4\nb\\
 &&+{\epsilon_\star} _2 {\epsilon_\star}_1^3 \left(8 \ln3-\frac{404}{27}\right)+{\epsilon_\star} _2 {\epsilon_\star} _3{\epsilon_\star} _1^2 \left(4 \ln3-\frac{40}{27}\right)\Big\}+\mathcal{O}({\epsilon_\star}_i^3).
 \eqn
The running of the scalar spectral index can be rewritten as
\bqn
\alpha_s \simeq \frac{1}{\left(2 {\epsilon_\star} _1-{\delta_\star} _1\right)^2} &&\Big(8 {\delta_\star} _1 {\epsilon_\star} _2 {\epsilon_\star} _1^2+2 {\delta_\star} _1 {\delta_\star} _2^2 {\epsilon_\star} _1+2 {\delta_\star} _1 {\epsilon_\star} _2^2 {\epsilon_\star} _1+2 {\delta_\star} _1 {\delta_\star} _2 {\delta_\star} _3 {\epsilon_\star} _1\nb\\
&& -2 {\delta_\star} _1^2 {\epsilon_\star} _2 {\epsilon_\star} _1-4 {\delta_\star} _1 {\delta_\star} _2 {\epsilon_\star} _2 {\epsilon_\star} _1+2 {\delta_\star} _1 {\epsilon_\star} _2 {\epsilon_\star} _3 {\epsilon_\star} _1\nb\\
&&-{\delta_\star} _1^2 {\delta_\star} _2 {\delta_\star} _3-8 {\epsilon_\star} _2 {\epsilon_\star} _1^3-4 {\epsilon_\star} _2 {\epsilon_\star} _3 {\epsilon_\star} _1^2\Big)+\mathcal{O}({\epsilon_\star}_i^3).
\eqn

Now we consider the tensor power spectrum, we obtain
\bqn
\Delta^2_h(k) &\simeq& \frac{181 H_{\star}^2}{36 e^3 \pi^2} \Big[1-\frac{{\delta_\star} _1}{2}+\left(-\frac{496}{181}+2\ln3\right) {\epsilon_\star} _1  -\frac{{\delta_\star} _1^2}{8}+\left(\frac{67}{181}+\frac{\ln3}{2}\right) {\delta_\star}
   _1 {\delta_\star} _2+\left(\frac{677}{362}-\ln3\right) {\delta_\star} _1 {\epsilon_\star}
   _1\nb \\
   &&+\left(\frac{517}{543}-\frac{630 \ln3}{181}+2 \ln^2{3}\right)
   {\epsilon_\star} _1^2+\left(-\frac{4636}{1629}+\frac{\pi ^2}{12}+\frac{496 \ln
   3}{181}-\ln^2{3}\right) {\epsilon_\star} _1 {\epsilon_\star} _2 +\mathcal{O}({\epsilon_\star}_i^3)\Big].
\eqn
The spectral index $n_h$ is
\bqn
n_t \simeq -2 {\epsilon_\star} _1-2 {\epsilon_\star} _1^2-\frac{{\delta_\star} _1 {\delta_\star} _2}{2}+\frac{1}{27}
   (-74+27 2\ln3) {\epsilon_\star} _1 {\epsilon_\star} _2+\mathcal{O}({\epsilon_\star}_i^3),
\eqn
and the running of the tensor spectral index reads
\bqn
\alpha_t &\simeq& -2 {\epsilon_\star} _1 {\epsilon_\star} _2+\mathcal{O}({\epsilon_\star}_i^3).
\eqn

With both the scalar and tensor spectra given, then the tensor-to-scalar ratio can be calculated as
\bqn
r &\equiv& \frac{8 \Delta_h(k)}{\Delta_{\mathcal{R}}(k)} =  8\Big|2 {\epsilon_\star} _1-{\delta_\star} _1+{\delta_\star} _1^2+\left(\frac{114}{181}+\ln
   3\right) {\delta_\star} _1 {\delta_\star} _2+\left(\frac{134}{181}-2 \ln
   3\right) {\epsilon_\star} _1 {\epsilon_\star} _2\Big|+\mathcal{O}({\epsilon_\star}_i^3).
\eqn

\subsubsection{$c_s(\eta_\star)<c_t(\eta_\star)$}

Then, similarly for $c_s(\eta_\star)<c_t(\eta_\star)$, we have to evaluate all the expression at the time when tensor mode crosses the Hubble horizon. Thus, we find that the scalar power spectrum,
\bqn
\Delta_{\mathcal{R}}^2(k) &=&\frac{181 H_{\star}^2}{36 e^3 \pi^2 (2 {\epsilon_\star} _1-{\delta_\star} _1)}\Big\{1+2\ln3{\epsilon_\star} _1+ \frac{1}{(2 {\epsilon_\star} _1-{\delta_\star} _1)}\Big[-\frac{{\delta_\star} _1^2}{2}-\left(\frac{114}{181}+\ln3\right) {\delta_\star} _1 {\delta_\star}
   _2\nb\\
   && +\frac{315}{181} {\delta_\star} _1 {\epsilon_\star}
   _1-\frac{992}{181} {\epsilon_\star}
   _1^2-\left(\frac{134}{181}-2 \ln3\right) {\epsilon_\star} _1 {\epsilon_\star} _2\Big] +\frac{1}{\left(2 {\epsilon_\star} _1-{\delta_\star} _1\right){}^2} \Big[{\delta_\star} _1^2 {\epsilon_\star}
   _1^2 \left(-\frac{569}{543}+2 \ln^2{3}-\frac{630 \ln
   3}{181}\right)\nb \\
   &&+{\delta_\star} _2 {\delta_\star} _1^2 {\epsilon_\star} _1 \left(-\frac{2318}{543}+2
   \ln^2{3}+\frac{456 \ln3}{181}\right)+{\delta_\star} _1^2 {\epsilon_\star} _1 {\epsilon_\star}
   _2 \left(-\frac{4456}{1629}+\frac{\pi ^2}{12}-\ln^2{3}-\frac{47 \ln
   3}{181}\right)\nb \\
   &&+{\delta_\star} _1 {\epsilon_\star} _1^3 \left(\frac{3080}{543}-8 \ln
   ^2{3}+\frac{1796 \ln3}{181}\right)+{\delta_\star} _2 {\delta_\star} _1 {\epsilon_\star} _1^2
   \left(\frac{1942}{543}-4 \ln^2{3}+\frac{536 \ln3}{181}\right)\nb \\
   &&+{\delta_\star}_1 {\epsilon_\star} _1 {\epsilon_\star} _2^2 \left(\frac{172}{1629}-\frac{\pi ^2}{12}+\ln
   ^2{3}-\frac{134 \ln3}{181}\right)+{\delta_\star} _2^2 {\delta_\star} _1 {\epsilon_\star} _1
   \left(-\frac{1034}{1629}-\frac{\pi ^2}{12}+\ln^2{3}+\frac{228 \ln
   3}{181}\right)\nb \\
   &&+{\delta_\star} _2 {\delta_\star} _3 {\delta_\star} _1 {\epsilon_\star} _1
   \left(-\frac{1034}{1629}-\frac{\pi ^2}{12}+\ln^2{3}+\frac{228 \ln
   3}{181}\right)+{\delta_\star} _2 {\delta_\star} _1 {\epsilon_\star} _1 {\epsilon_\star} _2
   \left(\frac{94}{1629}+\frac{\pi ^2}{6}-4 \ln^2{3}-\frac{188 \ln
   3}{181}\right)\nb \\
   &&+{\delta_\star} _1 {\epsilon_\star} _1 {\epsilon_\star} _2 {\epsilon_\star} _3
   \left(\frac{172}{1629}-\frac{\pi ^2}{12}+\ln^2{3}-\frac{134 \ln
   3}{181}\right)\nb\\
   && +{\delta_\star} _1^3 {\epsilon_\star} _1 \left(\ln
   3-\frac{315}{362}\right)+{\delta_\star} _1 {\epsilon_\star} _1^2 {\epsilon_\star} _2
   \left(\frac{21112}{1629}-\frac{\pi ^2}{3}-8 \ln3\right)\nb \\
   &&+\frac{3 {\delta_\star}
   _1^4}{4}+\frac{5}{2} {\delta_\star} _2 {\delta_\star} _1^3+{\delta_\star} _2^2 {\delta_\star} _1^2
   \left(\frac{1013}{1086}+\frac{\ln^2{3}}{2}+\frac{114 \ln
   3}{181}\right)+{\delta_\star} _2 {\delta_\star} _3 {\delta_\star} _1^2
   \left(\frac{517}{1629}+\frac{\pi ^2}{24}-\frac{\ln^2{3}}{2}-\frac{114 \ln
   3}{181}\right)\nb \\
   &&+{\epsilon_\star} _1^4 \left(\frac{2068}{543}+8 \ln
   ^2{3}-\frac{2520 \ln3}{181}\right)+{\epsilon_\star} _1^2 {\epsilon_\star} _2^2
   \left(\frac{658}{543}+2 \ln^2{3}-\frac{268 \ln3}{181}\right)\nb \\
   &&+{\epsilon_\star}
   _1^3 {\epsilon_\star} _2 \left(-\frac{14752}{1629}+\frac{\pi ^2}{3}+4 \ln
   ^2{3}+\frac{188 \ln3}{181}\right)\nb\\
   && +{\epsilon_\star} _1^2 {\epsilon_\star} _2 {\epsilon_\star} _3
   \left(-\frac{344}{1629}+\frac{\pi ^2}{6}-2 \ln^2{3}+\frac{268 \ln
   3}{181}\right)\Big]+\mathcal{O}({\epsilon_\star}_i^3),
   \Big\}
\eqn
the spectral index of scalar spectrum,
\bqn
n_s-1 &\simeq&-2{\epsilon_\star}_1+\frac{1}{2 {\epsilon_\star} _1-{\delta_\star} _1} \left({\delta_\star} _1 {\delta_\star} _2-2 {\epsilon_\star} _2 {\epsilon_\star} _1\right)\nb\\
&& +\frac{1}{\left(2 {\epsilon_\star} _1-{\delta_\star} _1\right){}^2}\Big[8 {\delta_\star} _1 {\epsilon_\star}
   _1^3-2 {\delta_\star} _1^2 {\epsilon_\star} _1^2-3 {\delta_\star} _1^2 {\delta_\star} _2 {\epsilon_\star} _1+{\delta_\star}
   _1 {\epsilon_\star} _2 {\epsilon_\star} _1^2 \left(\frac{350}{27}-8 \ln3\right)\nb\\
   &&+{\delta_\star}
   _1 {\delta_\star} _2^2 {\epsilon_\star} _1 \left(-\frac{34}{27}-2 \ln3\right)+{\delta_\star} _1
   {\epsilon_\star} _2^2 {\epsilon_\star} _1 \left(\frac{20}{27}-2 \ln3\right)\nb\\
   && +{\delta_\star} _1
   {\delta_\star} _2 {\delta_\star} _3 {\epsilon_\star} _1 \left(-\frac{34}{27}-2 \ln
   3\right)+{\delta_\star} _1^2 {\epsilon_\star} _2 {\epsilon_\star} _1 \left(2 \ln
   3-\frac{20}{27}\right)\nb\\
   &&+{\delta_\star} _1 {\delta_\star} _2 {\epsilon_\star} _2 {\epsilon_\star} _1
   \left(\frac{14}{27}+4 \ln3\right)+{\delta_\star} _1 {\epsilon_\star} _2 {\epsilon_\star} _3
   {\epsilon_\star} _1 \left(\frac{20}{27}-2 \ln3\right)+\frac{1}{2} {\delta_\star} _1^3
   {\delta_\star} _2+{\delta_\star} _1^2 {\delta_\star} _2 {\delta_\star} _3 \left(\frac{17}{27}+\ln
   3\right)-8 {\epsilon_\star} _1^4\nb\\
   &&+{\epsilon_\star} _2 {\epsilon_\star} _1^3 \left(8 \ln
   3-\frac{404}{27}\right)+{\epsilon_\star} _2 {\epsilon_\star} _3 {\epsilon_\star} _1^2 \left(4 \ln
   3-\frac{40}{27}\right)\Big]+\mathcal{O}({\epsilon_\star}_i^3),
\eqn
and the running of the scalar spectral index
\bqn
\alpha_s \simeq \frac{1}{\left(2 {\epsilon_\star} _1-{\delta_\star} _1\right)^2}
&&\Big(8 {\delta_\star} _1 {\epsilon_\star} _2 {\epsilon_\star} _1^2+2 {\delta_\star} _1 {\delta_\star} _2^2 {\epsilon_\star} _1+2 {\delta_\star} _1 {\epsilon_\star} _2^2 {\epsilon_\star} _1\nb\\
&& +2 {\delta_\star} _1 {\delta_\star} _2 {\delta_\star} _3 {\epsilon_\star} _1-2 {\delta_\star} _1^2 {\epsilon_\star} _2 {\epsilon_\star} _1-4 {\delta_\star} _1 {\delta_\star} _2 {\epsilon_\star} _2 {\epsilon_\star} _1+2 {\delta_\star} _1 {\epsilon_\star} _2 {\epsilon_\star} _3 {\epsilon_\star} _1\nb\\
&&-{\delta_\star} _1^2 {\delta_\star} _2 {\delta_\star} _3-8 {\epsilon_\star} _2 {\epsilon_\star} _1^3-4 {\epsilon_\star} _2 {\epsilon_\star} _3 {\epsilon_\star} _1^2\Big)+\mathcal{O}({\epsilon_\star}_i^3).
\eqn

For tensor perturbation we find the tensor spectrum
\bqn
\Delta^2_h (k)&\simeq& \frac{181 H_{\star}^2}{36 e^3 \pi ^2} \Big[1-\frac{{\delta_\star}
   _1}{2}+\left(-\frac{496}{181}+2 \ln3\right) {\epsilon_\star} _1-\frac{{\delta_\star}
   _1^2}{8}+\left(\frac{67}{181}+\frac{\ln3}{2}\right) {\delta_\star} _1 {\delta_\star}
   _2+\left(\frac{1039}{362}-\ln3\right) {\delta_\star} _1 {\epsilon_\star}
   _1\nb \\
   &&+\left(\frac{517}{543}-\frac{630 \ln3}{181}+2 \ln^2{3}\right)
   {\epsilon_\star} _1^2+\left(-\frac{4636}{1629}+\frac{\pi ^2}{12}+\frac{496 \ln
   {3}}{181}-\ln^2{3}\right) {\epsilon_\star} _1 {\epsilon_\star} _2+\mathcal{O}({\epsilon_\star}_i^3)\Big],
\eqn
the spectral index for the tensor spectrum
\bqn
n_t &\simeq&-2 {\epsilon_\star} _1-2 {\epsilon_\star} _1^2-\frac{{\delta_\star} _1 {\delta_\star} _2}{2}+\
   (-\frac{74}{27}+ 2\ln3) {\epsilon_\star} _1 {\epsilon_\star} _2+\mathcal{O}({\epsilon_\star}_i^3),
\eqn
and the running of the tensor spectral index
\bqn
\alpha_t &\simeq& -2 {\epsilon_\star} _1 {\epsilon_\star} _2+\mathcal{O}({\epsilon_\star}_i^3)
\eqn

Then the tensor-to-scalar ratio reads
\bqn
r &\simeq& 8\Big|2 {\epsilon_\star} _1-{\delta_\star} _1+{\delta_\star} _1^2+\left(\frac{114}{181}+\ln
   {3}\right) {\delta_\star} _1 {\delta_\star} _2+\left(\frac{134}{181}-2 \ln
   {3}\right) {\epsilon_\star} _1 {\epsilon_\star} _2\Big|+\mathcal{O}({\epsilon_\star}_i^3).
\eqn

\end{widetext}

\section{Conclusions and Outlook}

The uniform asymptotic approximation method provides a powerful, systematically improvable, and error-controlled approach to construct accurate analytical solutions of linear perturbations. Its effectiveness has been verified by applying it to the inflation models with nonlinear dispersion relations \cite{zhu_inflationary_2014, zhu_high-order_2016, zhu_gravitational_2014-2}, $k$-inflation \cite{zhu_power_2014}, and quantum corrections in loop quantum cosmology \cite{zhu_detecting_2015, zhu_scalar_2015, zhu_inflationary_2016}. In this paper, we applied the third-order uniform asymptotic approximation to derive the inflationary observables for scalar and tensor perturbations in the slow-roll inflation with the GB corrections. We obtained explicitly the analytical expressions of the power spectra, spectral indices, and running of spectral indices for both scalar and tensor perturbations in terms of the flow of the Hubble and GB coupling slow-roll parameters up to the second-order in the expansions of the slow-roll approximation. These results represent a significant improvement over the previous results obtained so far in the literature.

We would like to note that the results presented in this paper can be extended in several directions. First, it would be interesting to constrain the GB coupling function $\xi(\phi)$ and its derivatives by using the more precise forthcoming observational data. We expect that such constraints could help us to understand the physics of the early universe. Second, in our derivations of the primordial spectra, we assumed that the {\em time-dependent sound speed} in the slow-roll background is also slowly varying. It is also interesting to see what we will happen  if we relax this condition to allow the sound speed rapidly changing during a certain period. Finally, the method and calculations in this paper can be easily extended to various inflationary models with other types of higher-order curvature corrections. For example, in the Horndenski theories \cite{horndeski_second-order_1974, kobayashi_generalized_2011}, similar to the inflation with the GB coupling, both the inflationary scalar and tensor perturbations obey equations of motion that associated with time-dependent sound speeds. Thus, it is very interesting to see how the higher-order curvature terms in these theories affect  the primordial spectra in the slow-roll inflation. We hope to come back to these issues soon.

\section*{Acknowledgements}
This work is supported in part by National Natural Science Foundation of China with the Grants Nos. 11375153 (A.W.), 11675145(A.W.), 11675143 (T.Z. \& Q.W.), and 11105120 (T.Z.).

\end{document}